\documentclass[10pt,conference]{IEEEtran}
\usepackage{color,amsmath,amssymb,epsfig,multirow}
\begin{document}
\title{Sensing for Spectrum Sharing in Cognitive LTE-A Cellular Networks\vspace{-10mm}}
\author{\authorblockN{Prasanth Karunakaran\authorrefmark{1}, Thomas Wagner\authorrefmark{2}, Ansgar Scherb\authorrefmark{2} and Wolfgang Gerstacker\authorrefmark{1}} \authorblockA{\authorrefmark{1}Institute for Digital Communications\\Friedrich-Alexander-University Erlangen-N\"{u}rnberg (FAU), Germany\\Email: \{karunakaran,gersta\}@lnt.de}\authorblockA{\authorrefmark{2}Ericsson, N\"{u}rnberg, Germany\\Email: \{thomas.wagner,ansgar.scherb\}@stericsson.com}}

\maketitle

\begin{abstract}
In this work we present a case for dynamic spectrum sharing between different operators in systems with carrier aggregation (CA) which is an important feature in 3GPP LTE-A systems. Cross-carrier scheduling and sensing are identified as key enablers for such spectrum sharing in LTE-A. Sensing is classified as Type 1 sensing and Type 2 sensing and the role of each in the system operation is discussed. The more challenging Type 2 sensing which involves sensing the interfering signal in the presence of a desired signal is studied for a single-input single-output system. Energy detection and the most powerful test are formulated. The probability of false alarm and of detection are analyzed for energy detectors. Performance evaluations show that reasonable sensing performance can be achieved with the use of channel state information, making such sensing practically viable.
\end{abstract}

\makeatletter{\renewcommand*{\@makefnmark}{}
\footnotetext{This work has been performed in the framework of Cognitive Mobile Radio (CoMoRa) project, which is partly funded by the Federal Ministry of Education and Research (BMBF) of Germany.}\makeatother}
\vspace{-5mm}
%{\renewcommand{\thefootnote}{}
%{\footnotetext{This work has been performed in the framework of Cognitive Mobile Radio (CoMoRa) project, which is partly funded by the Federal Ministry of Education and Research (BMBF) of Germany.}}
\section{\vspace{-1mm}INTRODUCTION }
Dynamic spectrum access (DSA) based on cognitive radio technology is considered to be vital to achieve efficient spectrum utilization \cite{DySPA2012Mag}. In such systems, users are classified as either primary users or secondary users, where the primary users are the licensed users with highest priority. The secondary users sense the spectrum for primary user's presence and use the spectrum for transmission when it is free or transmit jointly with primary users without seriously affecting their communication. Mechanism designs for spectrum sharing in such cases are being developed \cite{AkyldizDyspan}. Such DSA can be considered for future cellular networks, especially in the context of spectrum sharing between different operators. Spectrum sharing allows a lightly loaded operator to dynamically share their spectrum to another operator thereby improving the spectrum utilization and the profitability. This idea becomes particularly relevant in the context of the 4G cellular standard 3GPP LTE-A due to the recently introduced carrier aggregation (CA) feature \cite{3gpp1}. CA allows an operator to aggregate up to five LTE-A carriers of 20 MHz bandwidth each. When a single operator exclusively owns a large bandwidth, the chance that the actual load from the coverage area being less than the capacity is higher. Hence a large spectrum deployment using CA is a suitable candidate for spectrum sharing.

In traditional cognitive radio (CR) primary users are unaware of the secondary users and it is the responsibility of the secondary user to ensure that it does not interfere with the primary user. Secondary users perform sensing to meet this requirement. Majority of the sensing studied in the CR literature determine whether the spectrum is vacant (i.e., usable) or not \cite{SensingSurvey}. This task belongs to the class of binary hypothesis testing problems where the competing hypotheses are the "noise" hypothesis ($H_0$) and the "signal + noise" hypothesis ($H_1$) \cite{SensingSurvey}\cite{SimonEnergyDetICC2003}. In such a case, the secondary device  perform sensing at the beginning of a slot and starts transmitting if the spectrum is sensed to be vacant. The main problem with such protocols is that it cannot react quickly if the primary transmission starts after the sensing interval causing interference till the end of the slot. Recently, this problem has been addressed in \cite{Wlee1}\cite{Robert1} where schemes have been developed that allow sensing during the transmission phase of the secondary transmitter. The competing hypotheses for this case are the "serving cell signal + noise" ($H_1^{'}$) hypothesis and the "serving cell signal + interfering signal + noise" hypothesis ($H_2$). In \cite{Wlee1} the authors propose zero-forcing beamforming by the secondary base station towards idle secondary users who can then sense as if there is no secondary signal. In \cite{Robert1} a multiple-input multiple-output (MIMO) system is considered with pre-processing and post-processing based on singular value decomposition (SVD) of channel matrix to decompose the system into parallel channels and the transmit power of each subchannel, sensing time in each subchannel and sensing threshold are optimized to maximize the CR throughput while constraining the probability of false alarm and of detection. These two approaches are not directly applicable to LTE-A systems, because the transmission modes and precoding are standardized and sensing must be evaluated with these restrictions. Therefore, in this work, we study the problem of sensing in presence of a signal for a single-input single-output configuration. MIMO configurations will be addressed in future works. 

Energy detection (ED) is one of the most important and well studied methods for testing between $H_0$ and $H_1$ \cite{SensingSurvey}\cite{SimonEnergyDetICC2003}. Even though ED is prone to noise variance uncertainties, it still remain the favored method for practical implementations due to its low computational complexity and ability to work with unknown signals. In this work, we adapt the energy detector of \cite{SimonEnergyDetICC2003} for distinguishing between $H_1^{'}$ and $H_2$. Then we extend the energy detector to incorporate channel knowledge, and the false alarm and detection probabilities are analyzed. Performance results show that meaningful detection and false alarm probabilities can be achieved if channel knowledge can be exploited. We also formulate the most powerful test for this scenario.   

The paper is organized as follows. Section \ref{sec:SharingModels} describes the spectrum sharing possibilities in 3GPP LTE/LTE-A networks, importance of cross-carrier scheduling, the role of sensing in spectrum sharing and the system model. In Section \ref{sec:ED} the energy detector of \cite{SimonEnergyDetICC2003} is formulated for our context and its extension to the case of availability of channel knowledge is provided. In Section \ref{sec:MPT} we state the most powerful test, and simulation results are presented in Section \ref{sec:PerfRes}. Section \ref{sec:Conclusion} provides the conclusion.

\section{\vspace{-1mm}Spectrum Sharing Models and Sensing}
\label{sec:SharingModels}

\begin{figure}[!t]
\centering
\includegraphics[width=8.5cm]{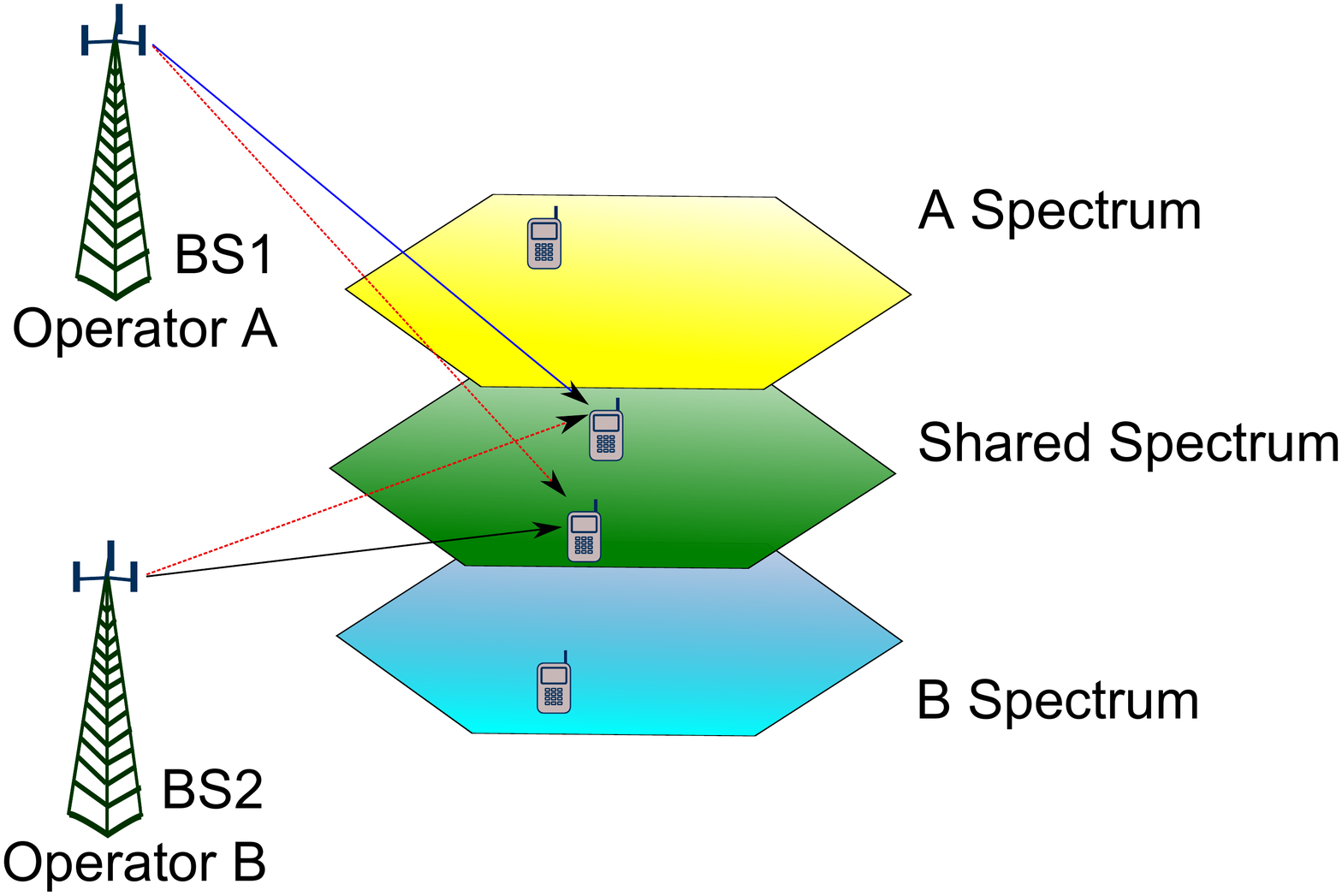}
\vspace{-5mm}
\caption{Spectrum sharing between two base stations}
\label{SystemFig1}
\end{figure}

Consider a spectrum sharing system composed of two 3GPP LTE base stations, BS1 and BS2, belonging to two different network operators as in Figure \ref{SystemFig1}. The operators are expected to have made certain agreements on spectrum sharing related to spectrum pricing, time scale of sharing, amount of tolerable interference caused to each other etc. Due to OFDMA transmission, LTE base stations can consider sharing with varying degrees of time-frequency granularity. The choices for spectrum sharing granularity in LTE can be understood from the frame structure of LTE \cite{3gpp1}. A radio frame in LTE occupies the full system bandwidth and spans $10$ ms. Each radio frame is composed of $10$ subframes, where each subframe is further divided in the frequency domain in units of physical resource blocks (PRBs). A PRB, which is also the smallest time-frequency scheduling granularity available in LTE systems, consists of $12$ contiguous subcarriers and $14$ OFDM symbols and occupies a $180$ kHz $\times$ $1$ ms slice in the time-frequency plane. Thus the granularity of spectrum sharing can be considered from the smallest unit of a PRB to the order of several radio frames.

Another restricting feature for spectrum sharing is the control channel structure. In LTE-A (Release 10) the Physical Downlink Control Channel (PDCCH) must be transmitted in the first three OFDM symbols of a subframe and are mapped across several subcarriers. If there is only one LTE-A carrier, then the control channels of both operators will collide and could adversely affect control channel coverage. In CA with multiple carriers, LTE-A supports scheduling transmissions over a carrier via control channels sent over another carrier. This feature, known as cross-carrier scheduling, allows to circumvent the control channel collisions if both operators have at least one exclusive carrier for themselves. The control channels can be sent over the exclusive carriers without collisions while spectrum sharing can happen in another aggregated carrier. Another possibility is the Enhanced-PDCCH (E-PDCCH) \cite{RSRel11} in the recently standardized version LTE-A Release 11. E-PDCCHs are not restricted to the first three OFDM symbols but transmitted in certain resources of the data channels and thus have the potential to be used in a spectrum shared operation.

We consider the downlink scenario and sensing is performed by user equipment (UE) (or dedicated sensing equipments installed by the operators). This is a reasonable assumption because it is the UE which is affected by downlink interference and the base station is usually not equipped with a downlink receiver. The sensing result could be reported back to the serving base station via uplink messaging. The sensing and reporting, in our view, is controlled by the base station. For example, the base station might instruct certain UEs to do sensing on particular PRBs or subframes. It is also assumed that the base station informs the selected UEs about whether the PRBs or subframes to be sensed carry the base station's transmissions or not. This assists the UE to determine which test to perform when the PRBs are not allocated for UE's own transmission. If the serving cell's signal is not present, the UE performs the traditional cognitive radio hypothesis test between $H_0$ and $H_1$. For the following discussions, it is assumed that BS1 has a higher priority than BS2. Based on the spectrum sharing granularity, we distinguish between two schemes.\vspace{-2.5mm}

\subsection{Long-Term Spectrum Sharing}
In long-term sharing, the spectrum sharing granularity is of the order of several hundreds of radio frames. Two possibilities can be considered for long-term sharing. In the first type, the spectrum sharing is controlled using licenses obtained via the backhaul. When BS1 does not require the spectrum, it turns off the transmission. BS2 can start using the spectrum when it is granted a license. Though sensing can aid BS2 in detecting the availability of spectrum, it is not essential for this type of operation. In the second type, sharing is controlled via sensing alone and does not involve license acquisition. This is more suitable for a relatively smaller granularity, say tens of radio frames. A performance criterion for sharing could be an upper limit on the degradation caused to the higher priority user. The lower priority operator could be charged a higher price if it violates the upper limit. As mentioned earlier, two types of sensing can be considered in this context. In Type 1 sensing, the goal is to distinguish between signal and noise. Type 1 sensing is mainly performed by the users of BS2 to detect the end of transmission by BS1. BS2 can then start to occupy the spectrum. The hypotheses for a single-input single-output system, involved in Type 1 sensing are given as
\vspace{-0.4mm}
\begin{equation}\label{LongTermtype1sensing}
  \begin{aligned}
    H_0 : y(kT_s) &= n(kT_s);\;&k = 1,2,...,N_t\\
    H_1 : y(kT_s) &= g(kT_s)+n(kT_s);\;&k = 1,2,...,N_t.
  \end{aligned}
\end{equation}where $y(t)$ is the received signal, $g(t)$ is the received signal from BS1 (or BS2 when sensing is performed by users of BS1), $n(t)$ is the additive white Gaussian noise (AWGN), $T_s$ is the receiver sampling interval and $N_t$ is the number of samples in a unit of spectrum sharing granularity. Energy detection, cyclostationarity based detection and other methods of detection of traditional cognitive radio systems can be used in Type 1 sensing. 

In Type 2 sensing, the sensing is performed while receiving the serving cell signal to detect the start of transmission by the other base station. Type 2 sensing is beneficial to both BS1 and BS2. For BS2, it allows to stop the transmission quickly when BS1 start its transmission again. Another advantage of Type 2 sensing for BS2 is that it allows another opportunity to rectify the miss-detection from Type 1 sensing. The use of Type 2 sensing for BS1 is to identify the interference from BS2 and the statistics of such events could be used to assess the impact of spectrum sharing with BS2. The hypotheses involved in Type 2 sensing are given by
\vspace{-0.4mm}
\begin{align}\label{LongTermtype2sensing}
  %\begin{aligned}
    H_1^{'} : y(kT_s) &= s(kT_s)+n(kT_s);~k = 1,...,N_t\\\nonumber
    H_2 : y(kT_s) &= s(kT_s)+g(kT_s)+n(kT_s);~k = 1,...,N_t.
  %\end{aligned}
\end{align}where $s(t)$ is the received signal from serving base station and the other parameters are same as in (\ref{LongTermtype1sensing}). Energy detection is possible for Type 2 sensing. However, cyclostationarity based detection is difficult as both $s(t)$and $g(t)$ have similar cyclostationary features.\vspace{-2.5mm}

\subsection{Short-Term Spectrum Sharing}
In short-term spectrum sharing, sharing can be performed within a subframe with the smallest granularity of a physical resource block (PRB). BS1 and BS2 cannot be turned off completely during a short interval because LTE base stations must transmit signals such as common reference signals (CRS), synchronization signals, control channels etc. Synchronization of the base stations and certain scheduling restrictions might be necessary to support short-term spectrum sharing. The basic operation is as follows. When BS2 wants to transmit in a particular PRB, it performs Type 1 sensing in that PRB in the current subframe to decide whether it can transmit data on that PRB in the next subframe. If the PRB is not occupied in the current subframe, BS2 assumes that it is free and starts to transmit in the next subframe. There is always the possibility that in the next subframe that PRB is occupied due to new scheduling at BS1, making the sensing result from the previous subframe invalid. However, under low load conditions such collisions may not be frequent. Also, due to the random geometry of users, not all the collisions will result in data decoding failure. In any case, such causality problems could be circumvented by sacrificing some scheduling flexibility, for example, a scheduling bias may be introduced to favour time continuity of occupied PRBs. Type 1 sensing can also be performed by BS1 to reduce collisions with BS2 whenever possible. For example, if sensing reveals that there is a sufficient number of PRBs unused by BS2, BS1 can meet its users' demand with those. The hypotheses involved are the same as in (\ref{LongTermtype1sensing}). However, here the signals used must be filtered to pass only the frequency contents of the PRB of interest. Similar to testing for hypotheses in (\ref{LongTermtype1sensing}) energy detection, cyclostationarity based detection etc. could be employed for this. Some of these sensing methods could also be performed after the fast Fourier transform (FFT) of an OFDM receiver. It is easy to see that energy detection is possible after the FFT whereas cyclostationarity based methods are not (because the cyclic prefix is removed prior to the FFT). Note that for long-term spectrum sharing, time domain sensing should be preferred as the granularity in frequency is the whole bandwidth, and also the energy of the cyclic prefix can be utilized.

The inter-cell interference coordination (ICIC) and the enhanced ICIC (eICIC) features of LTE-A could be made useful for spectrum sharing. However, these techniques were developed for handling the marco interference and the marco-pico interference suffered by the users belonging to a single operator. These schemes require partial scheduling information to be exchanged over the X2 interface between coordinating base stations. It is not clear whether such an information exchange is possible between different operators. Even though it is possible, sensing would supplement in achieving a higher flexibility as information exchange over X2 can happen only every 20 ms which alone contains 2000 PRBs in each 20 MHz carrier \cite{3gpp1}.

In the remainder of the paper, we describe Type 1 and Type 2 sensing from the point of view of a BS1 user. The hypotheses for sensing by a BS2 user can be easily written in a similar fashion. In Type 1 sensing, performed by a BS1 user, the competing hypotheses after the FFT (subcarrier domain) are given by
\vspace{-0.4mm}
\begin{equation}\label{eqn_Type1}
  \begin{aligned}
    H_0 : y_i &= {n_i};\;&i = 1,2,...,N\\
    H_1 : y_i &= {h_2}{x_{2i}}+{n_i};\;&i = 1,2,...,N,
  \end{aligned}
\end{equation}where $y_i$ is the received symbol in the $i$th resource element, $h_2~{\sim}~CN(0,{\sigma_2^2})$ is the complex Gaussian channel coefficient from BS2 (interfering cell) with variance $\sigma_2^2$, $n_i~{\sim}~CN(0,{\sigma_n^2})$ is the AWGN at the receiver, $x_{2i}$ is the transmit symbol from BS2 in the $i$th resource element of the PRB with symbol energy $E_x=E[|x_{2i}|^2]$, and $N$ represents the number of usable resource elements in a PRB. This model assumes Rayleigh fading and is accurate under low frequency and time selectivity conditions.

As in long-term sharing, Type 2 sensing could be performed by BS2 users to check if BS1 is transmitting on a PRB where BS2 is already active. This can help BS2 to control the interference generated due to either a miss-detection in Type 1 sensing or a scheduling change at BS1. Type 2 sensing also allows BS1 to identify interfering events caused by BS2. Cyclostationarity based detection is not possible for Type 2, but energy detection and other methods could be considered. For a BS1 user, the hypotheses of Type 2 sensing after the FFT are given by 
%\footnote{For long-term spectrum sharing, the subcarrier domain model can be easily applied with an appropriate increase in the value of $N$. However, it makes more sense to perform sensing in time domain as the frequency domain granularity is the whole bandwidth, and the energy of the cyclic prefix is not wasted.} 
\vspace{-0.4mm}
\begin{equation}\label{eqn_Type2}
  \begin{aligned}
    H_1^{'} : y_i &= {h_1}{x_{1i}}+{n_i};\;&i = 1,2,...,N\\
    H_2 : y_i &= {h_1}{x_{1i}}+{h_2}{x_{2i}}+{n_i};\;&i = 1,2,...,N,
  \end{aligned}
\end{equation}where $h_1~{\sim}~CN(0,{\sigma_1^2})$ is the complex Gaussian channel coefficient from BS1 (serving cell) with variance $\sigma_1^2$, $x_{1i}$ is the transmit symbol from BS1 in the $i$th resource element of the PRB with symbol energy $E_x=E[|x_{1i}|^2]$ and the other parameters same as in (\ref{eqn_Type1}).

%\footnotetext[1]{For long-term spectrum sharing, the subcarrier domain model can be easily applied with an appropriate increase in the value of $N$. However, it makes more sense to perform sensing in time domain as the frequency domain granularity is the whole bandwidth, and the energy of the cyclic prefix is not wasted.}

Algorithms for Type 2 sensing are developed in the following sections and the sensing/detection performance is characterized in terms of the probability of false alarm ($P_f$) and the probability of detection ($P_d$) \cite{KayDet} defined as 
  \begin{align}
    P_f &= P(\text{Decide }H_2|H_1^{'}\text{ True})\\
    P_d &= P(\text{Decide }H_2|H_2\text{ True}).
  \end{align}

\section {\vspace{-1mm}Energy Detection}
\label{sec:ED}

\subsection{\vspace{-1mm}Energy detector without channel knowledge (ED1)}
The energy detector without channel knowledge is the same as that of \cite{SimonEnergyDetICC2003}, but applied for differentiating between $H_1^{'}$ and $H_2$, and is given by
\begin{equation}\label{eqn_ed1}
\begin{aligned}
&e_1=\frac{2}{{\sigma_n^2}}\sum_{i=1}^{N}{|y_i|^2}\\
&\text{Decide }H_2\text{ if }e_1 > t_1.
\end{aligned}
\end{equation}

Given $h_1$, $h_2$, ${x_{1i}}$ and ${x_{2i}}$, the PDFs of $y_i$ under $H_1^{'}$ and $H_2$ are $CN(h_1{x_{1i}},{\sigma_n^2})$ and $CN(h_1{x_{1i}}+h_2{x_{2i}},{\sigma_n^2})$ respectively. Hence the PDF of $e_1$ under both hypotheses is a non-central chi-square PDF with $2N$ degrees of freedom but with different non-centrality parameters ${\lambda}_1$ and ${\lambda}_2$ [\cite{Proakis}, Eq. (2.1.117-124)].
\vspace{-0.5mm}
\begin{equation}\label{eqn_ed1_stat}
\begin{aligned}
H_1^{'}: e_1~{\sim}~{\chi}_{2N}^{2}({\lambda}_1)\\
\text{where }{{\lambda}_1}&=\frac{2}{{\sigma_n^2}}|h_1|^2\sum_{i=1}^N{|x_{1i}|^2}\\
&{\approx}\frac{2N}{{\sigma_n^2}}|h_1|^2E_{x}\\
H_2: e_1~{\sim}~{\chi}_{2N}^{2}({\lambda}_2)\\
\text{where }{{\lambda}_2}&=\frac{2}{{\sigma_n^2}} \sum_{i=1}^N{|h_1x_{1i}+h_2x_{2i}|^2}\\
&{\approx}\frac{2N}{{\sigma_n^2}}(|h_1|^2+|h_2|^2)E_x.
\end{aligned}
\end{equation}
\vspace{-0.5mm}
For a given channel state $h_1$, a false alarm occurs when the energy exceeds the threshold $t_1$ and the corresponding probability is given by\footnotemark[1] 
\begin{align}\label{eqn_pfpd_ed1}
P_f(t_1|h_1)&=P(e_1>t_1|h_1)\\\nonumber
&=Q_N(\sqrt{\lambda}_1,\sqrt{t_1})\\\nonumber
&=Q_N\left(\sqrt{\frac{2N}{{\sigma_n^2}}|h_1|^2E_{x}},\sqrt{t_1}\right),\\\nonumber
\end{align}where $Q_N(a,b)$ is the generalized Marcum Q-function \cite{SimonEnergyDetICC2003}\cite{Proakis}\cite{QMint} and the probability of false alarm in Rayleigh fading is computed by averaging over the channel power $\gamma_1=|h_1|^2$
\begin{equation}\label{eqn_ed1_pf1}
%\begin{aligned}
P_f(t_1)=\int_{0}^{\infty }Q_N\left(\sqrt{\frac{2N}{{\sigma_n^2}}\gamma _1E_{x}},\sqrt{t_1}\right)\frac{1}{\sigma_1^2}\text{e}^{\left(-\frac{\gamma_1}{\sigma_1^2}\right)}\text{d}\gamma_1.
%\end{aligned}
\end{equation}
As in \cite{SimonEnergyDetICC2003} we use [\cite{QMint}, Eq. (30)] with $p^2 = \frac{\sigma_n^2}{NE_x\sigma_1^2}$ and obtain the closed-form expression for $P_f(t_1)$
\begin{align}\label{pf_ed1_closedform}
%\begin{split}
P_f(t_1)&=\text{e}^{-\frac{t_1}{2}} \biggl\{\sum_{n=0}^{N-2}{\frac{1}{n!}}{\left ( \frac{t_1}{2} \right )^n} + \\\nonumber
&(p^2+1)^{N-1}\left [ \text{e}^{\frac{t_1}{2(p^2+1)}} - \sum_{n=0}^{N-2}{\frac{1}{n!}} \left ( \frac{t_1}{2(p^2+1)} \right )^n \right ]\biggr\}.
%\end{split}
\end{align}
\footnotetext[1]{For large $N$, the data randomness gets averaged out in (\ref{eqn_ed1_stat}) and the non-centrality parameters only depend on the average symbol energy $E_x$.}

Similarly, the probability of detection in Rayleigh fading is obtained by averaging over both channel powers $\gamma_1=|h_1|^2$ and $\gamma_2=|h_2|^2$
\begin{align}\label{eqn_ed1_pd}
P_d(t_1)=\int_{0}^{\infty }\int_{0}^{\infty }Q_N\left(\sqrt{\frac{2N}{{\sigma_n^2}}(\gamma_1+\gamma_2)E_x},\sqrt{t_1}\right)\times\\\nonumber
\frac{1}{\sigma_1^2}\text{e}^{\left(-\frac{\gamma_1}{\sigma_1^2}\right)}\frac{1}{\sigma_2^2}\text{e}^{\left(-\frac{\gamma_2}{\sigma_2^2}\right)}\text{d}\gamma_1\text{d}\gamma_2.
\end{align}
We were not able to obtain a closed-form solution for the above integral. If needed, lookup tables could be constructed from offline numerical integrations. 

\subsection{\vspace{-1mm}Energy detector with channel knowledge (ED2)}

Detectors are usually built to maintain a fixed average false alarm probability of $\delta_{P_f}$. $\delta_{P_f}$ is usually set to be small, for example 5\%. Then the objective is to maximize the probability of detection for the fixed false alarm probability (Neyman-Pearson criterion \cite{KayDet}\cite{DetTheoryCommMag}). For an energy detector under time varying channel conditions, keeping a fixed threshold is not optimum. The performance can be improved by adjusting the threshold such that at each channel realization the threshold exactly meets the target false alarm probability. For a given channel realization $h_1$, the threshold corresponding to $\delta_{P_f}$ is computed using the inverse of (\ref{eqn_pfpd_ed1}) as

\begin{align}\label{eqn_ed2_optthreshold1}
%\begin{aligned}
%&t(\lambda_1,\delta_{P_f})=Q_N^{-1}\left(\sqrt{\lambda_1},\delta_{P_f} \right)
\{t(\lambda_1,\delta_{P_f})=u\;{\vert}\;Q_N\left(\sqrt{\lambda_1},\sqrt{u}\right)=\delta_{P_f}\}.
%\end{aligned}
\end{align}

\vspace{-0.3mm}

The corresponding detector is given by
\begin{align}\label{eqn_ed2_optthreshold2}
%\begin{aligned}
&e_2=\frac{2}{{\sigma_n^2}}\sum_{i=1}^{N}{|y_i|^2}\\\nonumber
&\text{Decide }H_2\text{ if }e_2 > t(\lambda_1,\delta_{P_f}).
%\end{aligned}
\end{align}

However, in order to realize the detector of (\ref{eqn_ed2_optthreshold2}), the non centrality parameter $\lambda_1=\frac{2N}{{\sigma_n^2}}|h_1|^2E_{x}$ should be computed which in turn is a function of channel power $|h_1|^2$. Fortunately, this is possible in 3GPP LTE where the base station transmits the pilot signals known as common reference signals (CRS) always. These signals are wide band and are available in every PRB even if data transmission does not occur in that PRB. This means that estimation of serving cell channel $h_1$ can always be performed. The average probability of false alarm of the detector is $\delta_{P_f}$ because the threshold $t(\lambda_1,\delta_{P_f})$ achieves false alarm probability of $\delta_{P_f}$  for every channel state $h_1$. The probability of detection is obtained by replacing $t_1$ in (\ref{eqn_ed1_pd}) with $t(\lambda_1,\delta_{P_f})$. 

For a non-central chi square PDF, the tail probability at a particular value monotonically increases with non-centrality parameter. Hence the threshold $t(\lambda_1,\delta_{P_f})$ corresponding to a given tail probability of $\delta_{P_f}$ also monotonically increases with non-centrality parameter $\lambda_1$. Intuitively this makes sense because the threshold should be higher when the desired signal power is higher and vice versa. This motivates an approximation to the detector of (\ref{eqn_ed2_optthreshold2}) by varying the threshold linearly with desired signal power. Specifically we choose the threshold as 

\begin{align}\label{eqn_ed2_linearthreshold}
%\begin{aligned}
t(|h_1|^2)=t_2 + E[e_2|H_1^{'}]=t_2+\frac{2N}{{\sigma_n^2}}(|h_1|^2{E_x}+{\sigma_n^2}).
%\end{aligned}
\end{align}

Here $t_2$ is a control parameter to achieve the desired probability of false alarm. The advantage of such an approximation is that it does not involve frequent computations of the inverse of the generalized Marcum Q-function and hence is computationally simpler. The probability of false alarm and detection with linear threshold is obtained by replacing $t_1$ with $t(|h_1|^2)$ in (\ref{eqn_ed1_pf1}) and (\ref{eqn_ed1_pd}) respectively, making them functions of $t_2$ alone.

\section{\vspace{-1mm}Most Powerful Test}
\label{sec:MPT}

The most powerful test (MPT) maximizes the probability of detection for a fixed probability of false alarm \cite{KayDet}\cite{DetTheoryCommMag}. The test is based on the ratio of joint PDFs of samples $\boldsymbol{y}=(y_1,...,y_N)^T$ under $H_1^{'}$ and $H_2$ and is given by
%\begin{equation}
%\begin{aligned}
\begin{align}
&L=\frac{p(\boldsymbol{y}|H_2)}{p(\boldsymbol{y}|H_1^{'})}\\\nonumber
&\text{Decide }H_2\text{ if }L > t.
\end{align}
%\end{aligned}
%\end{equation}
Under $H_1^{'}$ and  a given modulation signal set $\mathcal{X}_{1k}$, the received samples $y_i$ have a Gaussian mixture PDF,
%\begin{equation}
%\begin{aligned}
\begin{align}
p(y_i|h_1,\mathcal{X}_{1k},H_1^{'}) &= \sum_{x_{1i}\in\mathcal{X}_{1k}}{P(x_{1i})f_{n}(y_i-h_1x_{1i})}\\\nonumber
&=\frac{1}{|\mathcal{X}_{1k}|}\sum_{x_{1i}\in\mathcal{X}_{1k}}{f_{n}(y_i-h_1x_{1i})},
\end{align}where $f_n(u)$ is the PDF of $n\sim CN(0,{\sigma_n^2})$ and $|\mathcal{X}_{1k}|$ represents the cardinality of $\mathcal{X}_{1k}$.
%\end{aligned}
%\end{equation}
Using the fact that in LTE the transmission format does not change within a PRB and assuming independent symbols, the joint probability density can be written as
\begin{equation}
p(\boldsymbol{y}|h_1,\mathcal{X}_{1k},H_1^{'})=\prod_{i=1}^{N}p(y_i|h_1,\mathcal{X}_{1k},H_1^{'}).
\end{equation}
%\vspace{0.3cm}
Typically UEs are only aware of modulation formats for its own data. Hence, when UEs are sensing PRBs which are not carrying its data, the modulation format is unknown. Thus the modulation format is unknown in general and the final joint PDF is obtained by averaging over all $M$ possible modulation formats
\begin{equation}
p(\boldsymbol{y}|H_1^{'})=\frac{1}{M}\sum_{k=1}^{M}\prod_{i=1}^{N}p(y_i|h_1,\mathcal{X}_{1k},H_1^{'}).
\end{equation}
Similarly, under $H_2$ we obtain
\begin{align}
p(y_i|h_1,h_2,\mathcal{X}_{1k},\mathcal{X}_{2l},H_2)&=\frac{1}{|\mathcal{X}_{1k}||\mathcal{X}_{2l}|}\times\\\nonumber
&\sum_{x_{1i}\in\mathcal{X}_{1k},x_{2i}\in\mathcal{X}_{2l}}\hspace{-8mm}{f_{n}(y_i-h_1x_{1i}-h_2x_{2i})}.
\end{align}
and 
\begin{align}
p(\boldsymbol{y}|H_2)=\frac{1}{M^2}\sum_{k=1}^{M}\sum_{l=1}^{M}\prod_{i=1}^{N}p(y_i|h_1,h_2,\mathcal{X}_{1k},\mathcal{X}_{2l},H_2).
\end{align}

For LTE systems, $M=3$ with $\mathcal{X}_{r1}=4$QAM, $\mathcal{X}_{r2}=16$QAM and $\mathcal{X}_{r3}=64$QAM and $r\in\{1,2\}$ for the system under investigation.

\section{\vspace{-1mm}Performance Results}
\label{sec:PerfRes}

Simulations have been carried out to evaluate the performance. All results are averaged over $10000$ channel realizations. In the figures, $\frac{{\sigma_1^2}}{{\sigma_2^2}}$ and $\frac{{\sigma_1^2}}{{\sigma_n^2}}$ identifies the signal-to-interference ratio and signal-to-noise ratio, respectively. 'ED2 (Exact)' identifies the energy detector with channel knowledge using the exact threshold (inverse of the generalized Marcum-Q function) and 'ED2 (Linear)' identifies the energy detector with linear threshold. The total number of resource elements in a PRB is $168$ ($12\times14$). However, not all of them can be used for sensing. This is because the common reference signals of the interfering base station are always on and do not indicate whether the resource block is used for data transmission. Without going into the precise details of the LTE standard, we assume a reference signal overhead of 15\% which gives a value of $N=142$. Simulations are performed under SINR conditions of ($\frac{{\sigma_1^2}}{{\sigma_2^2}}=0$ dB,$\frac{{\sigma_1^2}}{{\sigma_n^2}}=6$ dB) and ($\frac{{\sigma_1^2}}{{\sigma_2^2}}=6$ dB,$\frac{{\sigma_1^2}}{{\sigma_n^2}}=12$ dB). Similar SINR conditions are expected in cellular deployments.

\begin{figure}[!t]
\centering
\includegraphics[width=9cm]{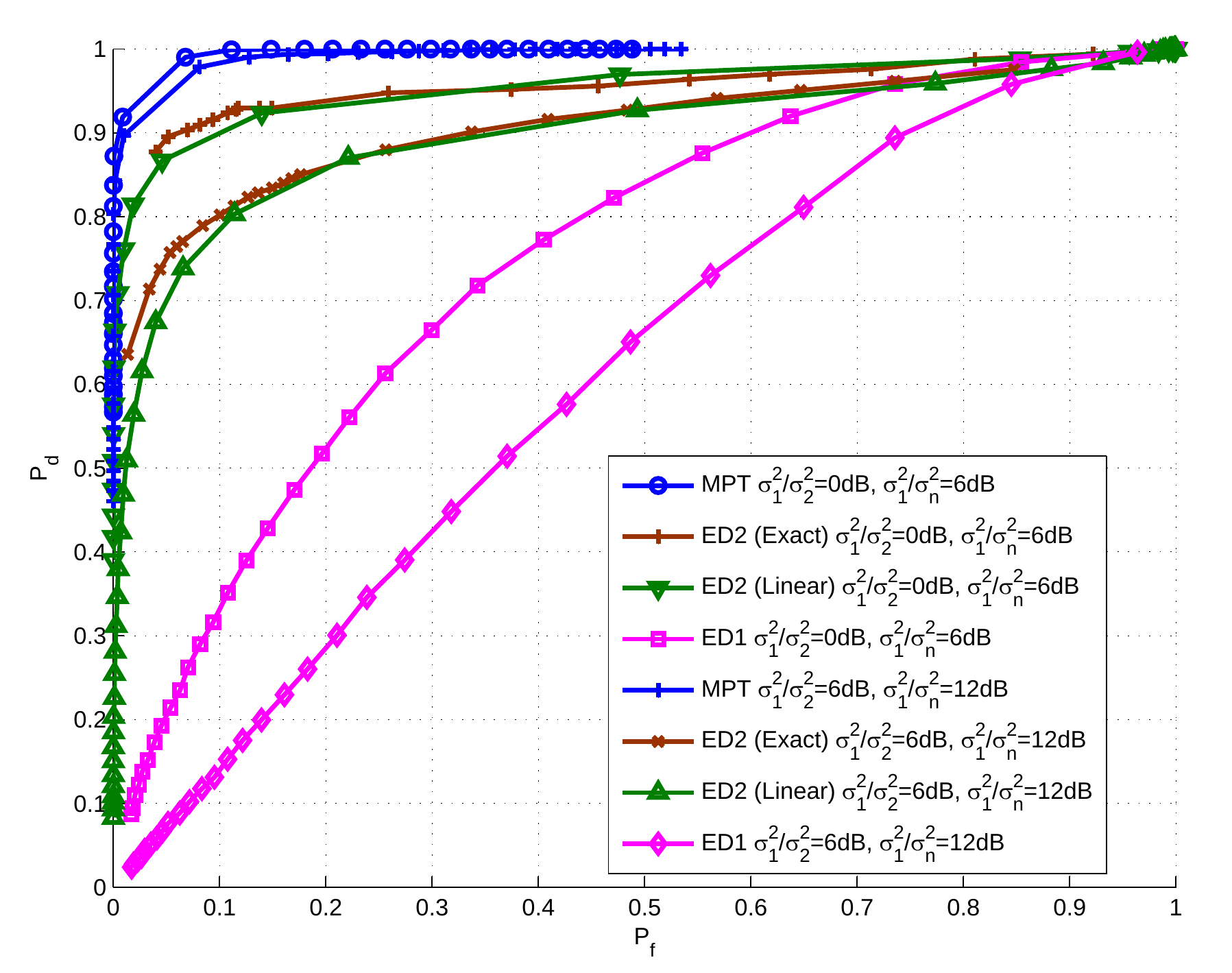}
\vspace{-3mm}
\caption{Receiver Operating Characteristics for ideal channel estimation.}
\label{RocIdealCh}
\end{figure}

The performance of detectors is captured by the receiver operation characteristics (ROC) curve which is the probability of detection depicted versus probability of false alarm \cite{KayDet}. Figure \ref{RocIdealCh} provides the ROC for the different detectors discussed earlier under ideal channel estimation. As expected the MPT performs best achieving greater than 90\% detection probability at a false alarm probability of 5\%. Even though such a detector is of prohibitive complexity, it serves as a performance benchmark \cite{KayDet}\cite{DetTheoryCommMag} for comparisons. The detection probabilities achieved by the energy detector without channel knowledge are quite low to have any real use. Reasonable detection probabilities are achieved, around 87\% at $\frac{{\sigma_1^2}}{{\sigma_2^2}}=0$ dB, by the energy detector with channel knowledge. Performance degrades as the interferer becomes weaker. The energy detector with linear threshold performs very close to the one with inverse generalized Marcum-Q threshold and indicates that the linear threshold is a suitable choice.

\begin{figure}[!t]
\centering
\includegraphics[width=8.5cm]{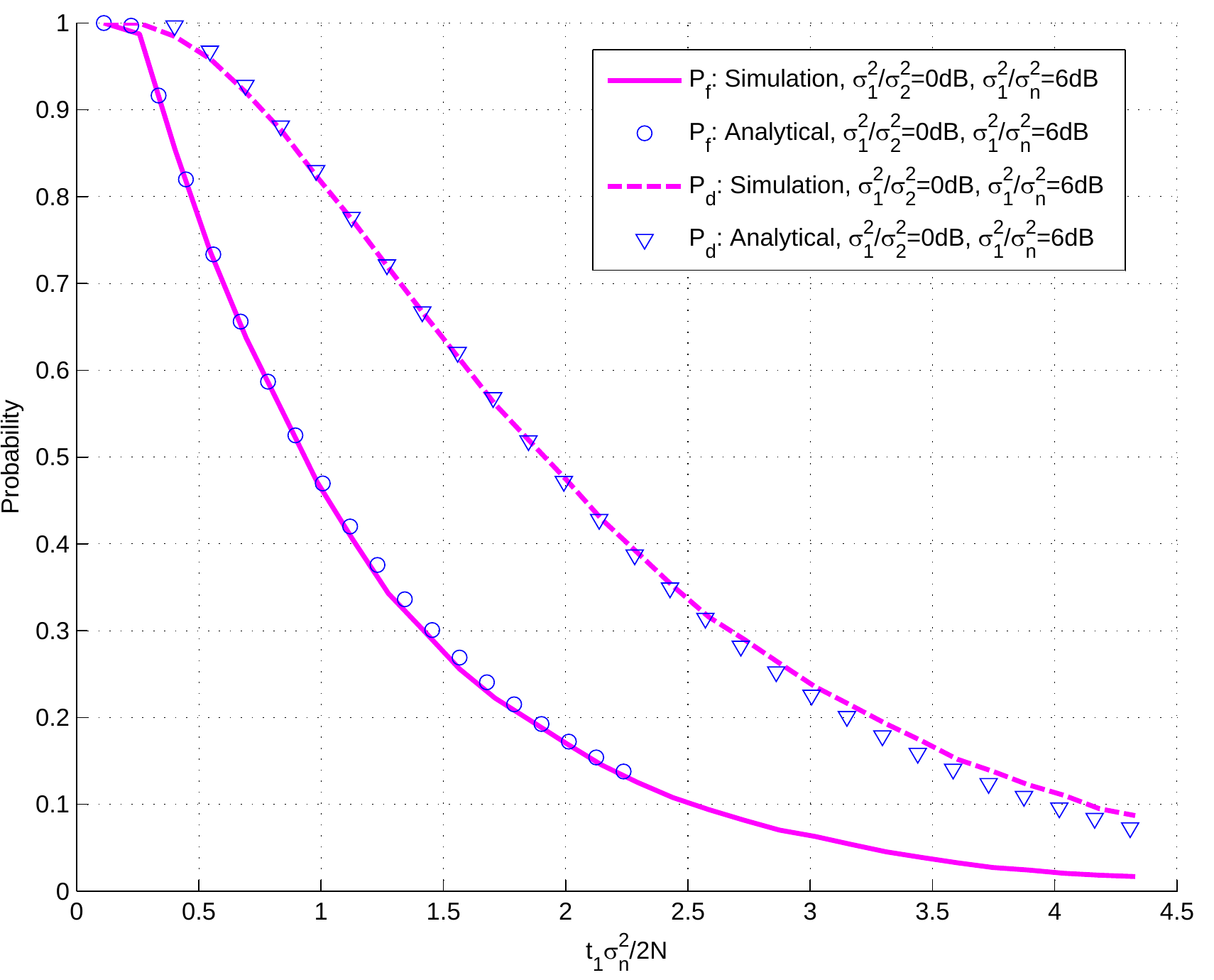}
\vspace{-3mm}
\caption{Comparison of analytical results and simulation results for ED1 with $N=142$.}
\label{pdpf_ed1}
\end{figure}

\begin{figure}[!t]
\centering
\includegraphics[width=8.5cm]{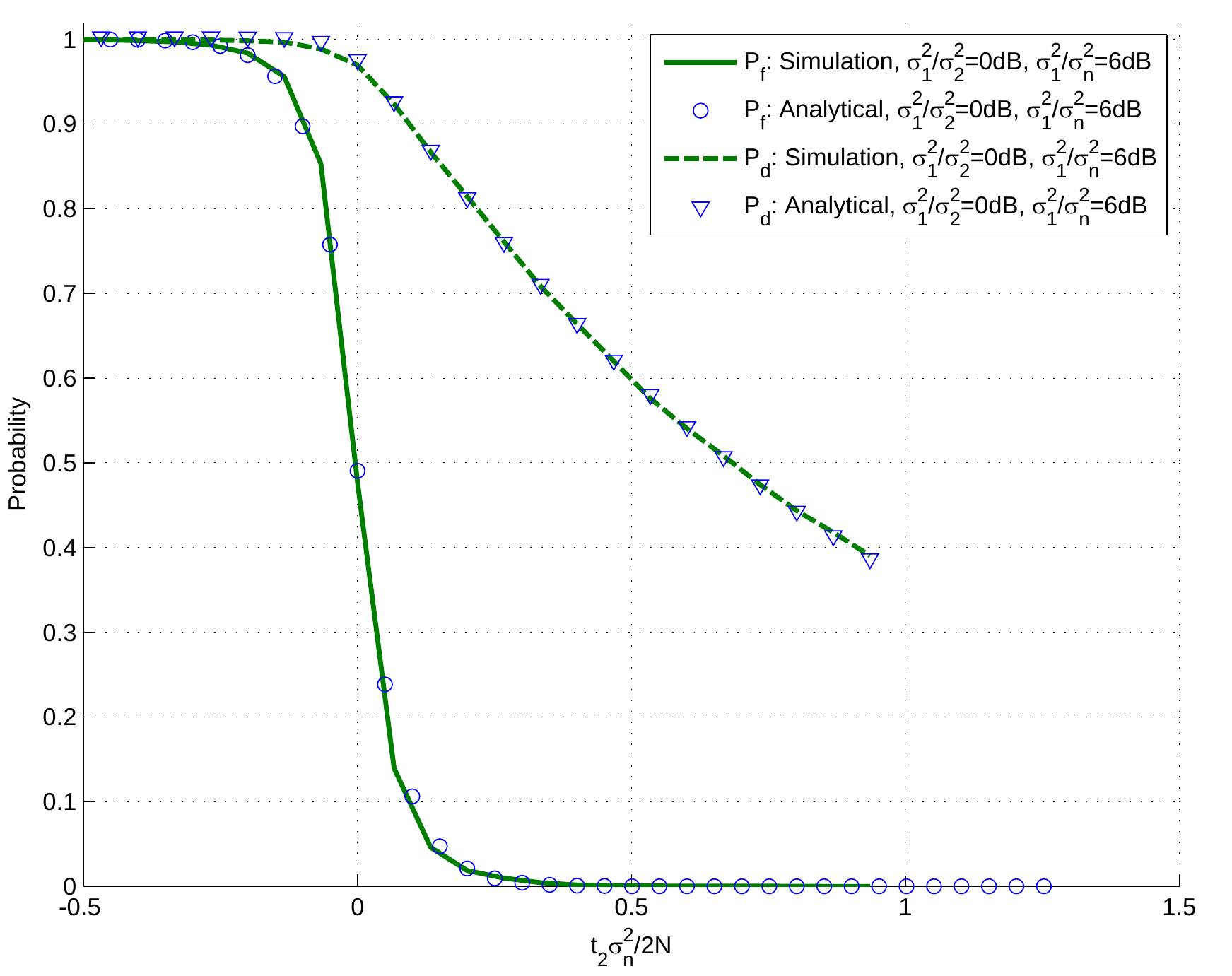}
\vspace{-3mm}
\caption{Comparison of analytical results and simulation results for ED2 (Linear) with $N=142$.}
\label{pdpf_ed2}
\end{figure}

Figures \ref{pdpf_ed1} and \ref{pdpf_ed2} compare the analytical and simulation results for the detector without channel knowledge and the detector with linear threshold, respectively. All analytical expressions except (\ref{pf_ed1_closedform}) were evaluated via numerical integration. The results shows a very close match between simulation and analytical results. Hence, analytical results can be used to set the operating threshold for the target false alarm probability.

\begin{figure}[!t]
\centering
\includegraphics[width=8.5cm]{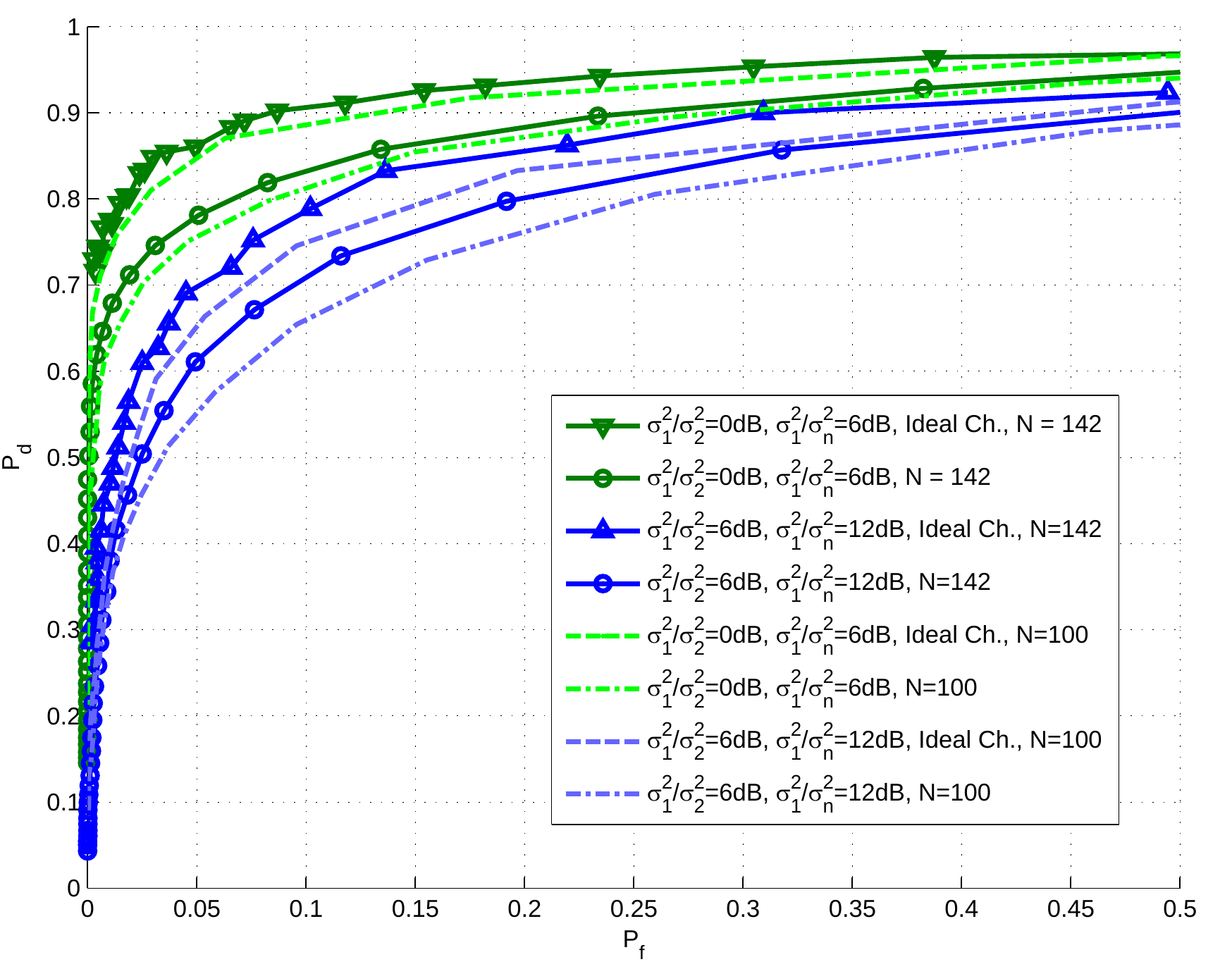}
\vspace{-3mm}
\caption{Effect of channel estimation error for ED2 (Linear).}
\label{cheesterr}
\end{figure}

In order to study the impact of channel estimation errors and the effect of $N$, we adopt an AWGN estimation error model. The error is introduced by injecting AWGN with variance equal to the mean square error of the estimator. The normalized mean square error is computed based on the CRS-LS curves of \cite{LTEChannEst} for ITU PedB and 0 km/h. Specifically, the normalized mean square error follows $\text{log}({\sigma_{\text{nmse}}^2})=-\frac{\text{SINR(dB)}}{10}-0.26$, for an SINR range of around 0 dB to 30 dB. The results are provided in Figure \ref{cheesterr}. A lower value of $N=100$ is also simulated. This corresponds to an overhead of $40\%$. Simulations show a decrease in detection probability of nearly $0.1$ (at $P_f=0.05$) due to estimation errors. Reducing the number of samples to 100 results in an additional drop of around $0.05$ in detection probability. The results show that reasonable detection probabilities can be obtained with the use of channel knowledge. Finally, we note that with the use of multiple antennas performance could be improved significantly. 

\vspace{-2.5mm}
\section{\vspace{-1mm}CONCLUSION}
\label{sec:Conclusion}

We have introduced the problem of sensing in presence of a desired signal which is relevant in the context of spectrum sharing between cellular operators. Detectors are developed and analyzed in the context of an LTE-A based system. It is shown that the performance of the energy detector can be significantly improved with the use of channel state information and achieves reasonable probability of detection. The results presented in the work provide also a lower bound for the performance with multiple antennas, which is expected to be further improved. These facts indicate that sensing in presence of a desired signal is a practically viable problem.

\end{document}